\def\msol{M_{\odot}}
\def\mbh{M_{\mathrm{BH}}}

\def\gr{$\gamma$-ray\ }

\def\grn{$\gamma$ ray\ }
\def\grnns{$\gamma$ ray}

\def\s4{S4 0954+65}
\def\fermi{\textit{Fermi}-LAT~}
\newcommand{\erg}{\mathrm{erg}}

\documentclass[longauth]{aa}  
\usepackage{amsmath}
\usepackage[switch]{lineno}
\usepackage{color}
\usepackage{graphicx}
\usepackage{txfonts}
\begin{document}

   \title{The detection of the blazar \s4 at very-high-energy with the MAGIC telescopes during an exceptionally high optical state}
 \titlerunning{\s4 February 2015 flare with MAGIC}

\author{
MAGIC Collaboration:
M.~L.~Ahnen\inst{1} \and
S.~Ansoldi\inst{2,20} \and
L.~A.~Antonelli\inst{3} \and
C.~Arcaro\inst{4} \and
D.~Baack\inst{5} \and
A.~Babi\'c\inst{6} \and
B.~Banerjee\inst{7} \and
P.~Bangale\inst{8} \and
U.~Barres de Almeida\inst{8,9} \and
J.~A.~Barrio\inst{10}
% ...
\and
J.A Acosta Pulido \inst{15,44}
}
\institute { ETH Zurich, CH-8093 Zurich, Switzerland
\and Universit\`a di Udine, and INFN Trieste, I-33100 Udine, Italy
\and National Institute for Astrophysics (INAF), I-00136 Rome, Italy
\and Universit\`a di Padova and INFN, I-35131 Padova, Italy
\and Technische Universit\"at Dortmund, D-44221 Dortmund, Germany
\and Croatian MAGIC Consortium: University of Rijeka, 51000 Rijeka, University of Split - FESB, 21000 Split,  University of Zagreb - FER, 10000 Zagreb, University of Osijek, 31000 Osijek and Rudjer Boskovic Institute, 10000 Zagreb, Croatia.
\and Saha Institute of Nuclear Physics, HBNI, 1/AF Bidhannagar, Salt Lake, Sector-1, Kolkata 700064, India
\and Max-Planck-Institut f\"ur Physik, D-80805 M\"unchen, Germany
\and now at Centro Brasileiro de Pesquisas F\'isicas (CBPF), 22290-180 URCA, Rio de Janeiro (RJ), Brasil
\and Unidad de Part\'iculas y Cosmolog\'ia (UPARCOS), Universidad Complutense, E-28040 Madrid, Spain
\and University of \L\'od\'z, Department of Astrophysics, PL-90236 \L\'od\'z, Poland
\and Deutsches Elektronen-Synchrotron (DESY), D-15738 Zeuthen, Germany
\and Institut de F\'isica d'Altes Energies (IFAE), The Barcelona Institute of Science and Technology (BIST), E-08193 Bellaterra (Barcelona), Spain
\and Universit\`a  di Siena and INFN Pisa, I-53100 Siena, Italy
\and Inst. de Astrof\'isica de Canarias, E-38200 La Laguna, and Universidad de La Laguna, Dpto. Astrof\'isica, E-38206 La Laguna, Tenerife, Spain
\and Universit\"at W\"urzburg, D-97074 W\"urzburg, Germany
\and Finnish MAGIC Consortium: Tuorla Observatory and Finnish Centre of Astronomy with ESO (FINCA), University of Turku, Vaisalantie 20, FI-21500 Piikki\"o, Astronomy Division, University of Oulu, FIN-90014 University of Oulu, Finland
\and Departament de F\'isica, and CERES-IEEC, Universitat Aut\'onoma de Barcelona, E-08193 Bellaterra, Spain
\and Universitat de Barcelona, ICC, IEEC-UB, E-08028 Barcelona, Spain
\and Japanese MAGIC Consortium: ICRR, The University of Tokyo, 277-8582 Chiba, Japan; Department of Physics, Kyoto University, 606-8502 Kyoto, Japan; Tokai University, 259-1292 Kanagawa, Japan; The University of Tokushima, 770-8502 Tokushima, Japan
\and Inst. for Nucl. Research and Nucl. Energy, Bulgarian Academy of Sciences, BG-1784 Sofia, Bulgaria
\and Universit\`a di Pisa, and INFN Pisa, I-56126 Pisa, Italy
\and Humboldt University of Berlin, Institut f\"ur Physik D-12489 Berlin Germany
\and also at Dipartimento di Fisica, Universit\`a di Trieste, I-34127 Trieste, Italy
\and also at Port d'Informaci\'o Cient\'ifica (PIC) E-08193 Bellaterra (Barcelona) Spain
\and also at INAF-Trieste and Dept. of Physics \& Astronomy, University of Bologna
\and NASA Goddard Space Flight Center, Greenbelt, MD 20771, USA and Department of Physics and Department of Astronomy, University of Maryland, College Park, MD 20742, USA
\and Hiroshima Astrophysical Science Center, Hiroshima, Japan
\and University of Maryland, Baltimore County, USA
\and The Catholic University of America, Washington DC, USA
\and Space Science Division, NRL, Washington DC, USA
\and Aalto University Metsahovi Radio Observatory, Finland
\and Aalto University Department of Electronics and Nanoengineering, Finland
\and Tartu Observatory, Estonia
\and Tuorla Observatory, University of Turku, V\"ais\"al\"antie 20, FI-21500 Piikki\"o, Finland
\and Astron.\ Inst., St. Petersburg State Univ., Russia 
\and Institute for Astrophysical Research, Boston University, USA
\and Pulkovo Observatory, St. Petersburg, Russia
\and Crimean Astrophysical Observatory, P/O Nauchny, Crimea, 298409, Russia
\and Instituto de Astrof\'{\i}sica de Andaluc\'{\i}a (CSIC), Apartado 3004, E--18080 Granada, Spain
\and Max--Planck--Institut f\"ur Radioastronomie, Auf dem H\"ugel, 69, D--53121, Bonn, Germany
\and Harvard-Smithsonian Center for Astrophysics, Cambridge, MA USA
\and INAF, Osservatorio Astrofisico di Torino, via Osservatorio 20, I-10025, Pino Torinese, Italy
\and Departamento de Astrofisica, Universidad de La Laguna, La Laguna, E-38205 Tenerife, Spain
}

   \date{Received ; accepted }
  \abstract
   {}   {The very-high-energy (VHE, $\gtrsim 100$ GeV) \gr MAGIC observations of the blazar \s4, were triggered by an exceptionally high flux state of emission in the optical.
    This blazar has a disputed redshift of $z$=0.368 or $z\geqslant$0.45 and an uncertain classification among blazar subclasses. The exceptional source state described here makes for an excellent opportunity to understand physical processes in the jet of \s4 and thus contribute to its classification.}
   {We investigate the multiwavelength (MWL) light curve and spectral energy distribution (SED) of the \s4 blazar during an enhanced state in February 2015 and put it in context with possible emission scenarios. We collect photometric data in radio, optical, X-ray, and \grnns. We study both the optical polarization and the inner parsec-scale jet behavior with 43 GHz data.}
   {Observations with the MAGIC telescopes led to the first detection of \s4 at VHE. Simultaneous data with \fermi at high energy \grn (HE, 100 MeV < E < 100 GeV) also show a period of increased activity. Imaging at 43 GHz reveals the emergence of a new feature in the radio jet in coincidence with the VHE flare. Simultaneous monitoring of the optical polarization angle reveals a rotation of approximately 100$^\circ$. }
   {The high emission state during the flare allows us to compile the simultaneous broadband SED and to characterize it in the scope of blazar jet emission models. The broadband spectrum can be modeled with an emission mechanism commonly invoked for flat spectrum radio quasars, i.e. inverse Compton scattering on an external soft photon field from the dust torus, also known as external Compton. The light curve and SED phenomenology is consistent with an interpretation of a blob propagating through a helical structured magnetic field and eventually crossing a standing shock in the jet, a scenario typically applied to flat spectrum radio quasars (FSRQs) and low-frequency peaked BL Lac objects (LBL).}
     \maketitle
%
\section{Introduction}
Blazars are a subclass of Active Galactic Nuclei (AGN) in which the relativistic jet presents a small viewing angle towards the observer and thus where relativistic effects on the observed emission are more extreme. Conventionally, blazars are subdivided in BL Lac objects and FSRQs depending on the characteristic of their optical spectrum: while BL Lac objects are dominated by the featureless continuum emission from the jet, FSRQs typically show wide optical emission lines. 

The blazar \s4 hosts a black hole of mass $\mbh\sim3.3\times 10^8 \msol$, estimated from the width of the H$_\alpha$ line \citep{blackholemasses}. The detection of the H$_\alpha$ line is not confirmed by \cite{landoni2015} (see the discussion on the redshift determination) so that the mass estimation cannot be confirmed either.
This blazar presents strong variability in the optical band, already well studied by \citet[][]{wagner1990}  and by \citet{morozova2014}. Intra night variability has been found both in optical and radio wavelengths \citep{wagner1993}.
The optical high brightness state of February 2015, presented here, is however exceptional for the object, with a brightening of more than 3 magnitudes in the R-band with respect to the average monitored state\footnote{\url{http://users.utu.fi/kani/1m/S4\_0954+65.html}}. This not only spurred many alerts in the community \citep[see ATel \#6996, \#7001, \#7057,~\#7083, \#7093;][]{atel6996,atel7001,atel7057,atel7083,atel7093}, but also the first and only detection of the object at very high energies (VHE, E$\gtrsim$100 GeV), thanks to observations by the MAGIC Telescopes. This detection by MAGIC and the MWL data collected alongside it are the focus of the present work.

The source GRO J0957+65, detected with the EGRET telescope on board the \textit{Compton Gamma-Ray Observatory}, has been associated through optical and radio observations with \s4 by \citet{egret95}. \s4 has been afterwards always included in the released catalogs of sources detected by the Large Area Telescope (LAT) instrument on board the \textit{Fermi} satellite \citep{1FGL,2FGL,1FHL,3FGL,2FHL,3FHL}, with the exclusion of the bright source list released after the first 3 months of \fermi data integration.

The classification of the object, based on the available literature, is still unclear. In most of the ATels mentioned above \s4 is referenced as a FSRQ, but in most of the literature this is classified as a BL Lac object due to the small equivalent width of the emission lines in its spectrum \citep[see, e.g.][]{stickel91}. \citet{sambruna1996} classified the SED of \s4 as ``FSRQ-like'', in a sample limited to the sources with a detection from EGRET data. It indeed presents a flatter spectral index than most BL Lac objects, in both X-ray and \gr bands \citep[see][ and references therein]{raiteri}. 
Among BL Lac objects, a further phenomenological subdivision can be made based on the frequency of the synchrotron peak, ranging from optical to X-ray frequency and identifying the classes of low-, intermediate- or high-peaked BL Lac object (LBL, IBL, HBL respectively). \citet{ghisellini2011} classified this object as a LBL based on the SED.
When including the kinematic features from the radio jet in the classification templates, \citet{hervetetal2016} classify this as their kinematic class II, mostly composed of FSRQ.
\s4 can thus be interpreted as a transitional object between FSRQ and classical BL Lac objects. 

The most numerous extragalactic sources detected at VHE from Imaging Air Cherenkov Telescopes (IACTs), presently, belong to the HBL class. Therefore the VHE detection of an object such as \s4 provides a rare opportunity to study VHE emission conceivably produced in a different kind of environment. Indeed, while emission in HBL can mostly be satisfactorily modeled taking into account only processes in a compact feature in the jet, for FSRQs the inclusion of the interactions of such a feature with the surrounding ambient becomes of greater importance \citep[see e.g. ][]{tavecchiogamma}. The structure of the broadband SED collected here will also be put in context with other common characteristics of a FSRQ classification, such as intrinsic brightness, peak of the synchrotron component and Compton dominance.

Also the question of \s4 redshift is still not settled, as claims of line detection in the optical spectrum are not always confirmed. The redshift of the source was first determined at $z$=0.368 by the identification of 
lines by \citet{law86,law96}. \citet{stickel93} obtained, from different measurements, the same redshift estimate based on line identification. None of these lines were confirmed by the observations reported in \citet{landoni2015}, who instead pose a lower limit of z$\geq0.45$. The latter results were obtained with a superior resolution spectra. At the time of the observation the magnitude in R-band of the object was 15.5, while it is known from variability studies that it could be even 2 magnitudes lower.  
In the following we will adopt the redshift $z$=0.368.

The outline of this paper is as follows. In Section \ref{sec:magicdata}, we will present the MAGIC telescopes and the relative data set on \s4. Section \ref{sec:mwlcoverage} reviews all the MWL data that were collected during this exceptional burst, whereas Section \ref{sec:lightcurveandsed} discusses the implication of this burst for the source state and inner jet structure. Additional information on the MAGIC data analysis, the parameters derived from the VLBA data and the full dataset for Swift-XRT X-ray data will be found in Appendix \ref{app:magicdata} and \ref{app:vlbadata}, respectively.


\section{MAGIC Observations}\label{sec:magicdata}
The MAGIC telescopes are an array of two IACTs located in the Island of La Palma (Spain) at an altitude of $\sim$ 2200 m asl. The system is sensitive down to an energy threshold of $E\sim50$ GeV \citep{magicperformancepaper} for low zenith angle observations. This is of particular relevance for the monitoring of variable sources and of those that tend to exhibit a steep spectrum at VHE. 
The full data have been analyzed using the standard MAGIC analysis chain and the MAGIC Standard Analysis Software  \citep[MARS,][]{standardMARS, magicperformancepaper}.

The MAGIC collaboration supports a program of Targets of Opportunity (ToO), triggered by MWL monitoring. The ToO program was activated for observations of \s4 at the end of January 2015 after the first hints of enhanced optical state (triggered by the Tuorla monitoring in R-band, see Section \ref{subsec:optical}).
We observed the source with the MAGIC telescopes for 2 nights (MJD 57049-57050, 2015 January 27 and 28), for a total of 1 hour high-quality dark time data, but obtained no detection. 
We resumed the ToO observations in February after the Tuorla monitoring revealed a very exceptional flux state, later confirmed by other monitoring programs (see Section \ref{subsec:optical}). We obtained a detection at a significance of $\sim7.4\sigma$ from observations during 2015 February 14 \citep[MJD 57067, ATel \#7080][]{atelmagic}.  
We continued observing \s4, barring adverse atmospheric conditions, until full moon days when standard MAGIC observations are not possible due to the elevated level of background light (last day of observation, with already large moonlight contamination, on 2015 March 1, MJD 57082). A detailed breakdown of the observation conditions and relative results can be found in Appendix \ref{app:magicdata}.

\begin{figure}
   \centering
   \includegraphics[width=\hsize]{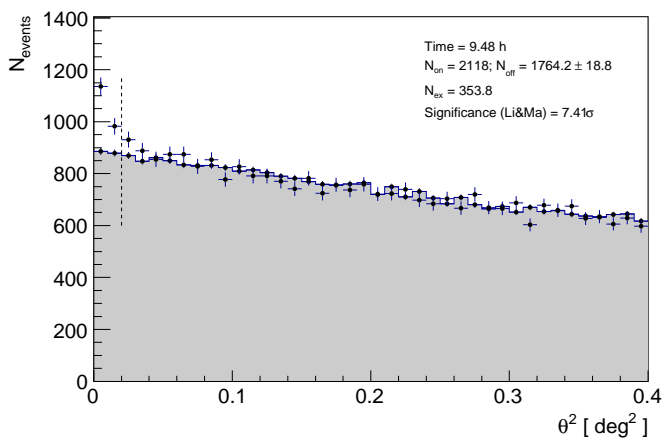}
      \caption{Distribution of the squared angular distance ($\theta^2$) between the reconstructed event direction and the nominal source  direction. The filled histogram is the background estimation, obtained from sky regions within the field of view with similar detector acceptance. We show only data taken in dark condition (condition 1, see Appendix \ref{app:magicdata}). The standard MAGIC low energy (LE) cuts are applied to the data (see Appendix \ref{app:magicdata} and Table \ref{table:magicdata}). The vertical line corresponds to the optimal cut ($\theta^2=0.02\mathrm{\ deg}^2$) for point source analysis in LE cuts, used to derive significance values.}
         \label{fig:odie_le_fullset}
   \end{figure}

The total excess from the dark-time data is consistent with a point source emission (see Fig. \ref{fig:odie_le_fullset}). No other significant emission is found in the field of view apart from the one coincident with \s4 at the center.

The SED points presented in Section \ref{sec:lightcurveandsed} below are derived for the day of the flare (MJD 57067, 2015 February 14), using only data taken in dark conditions (that allow for the lowest threshold and lowest systematic uncertainty, Appendix \ref{app:magicdata}). We follow the standard MAGIC unfolding procedure \citep{unfoldingmagic} to obtain the intrinsic spectrum. 

\begin{figure}[!htb]
    {\includegraphics[width=\linewidth]{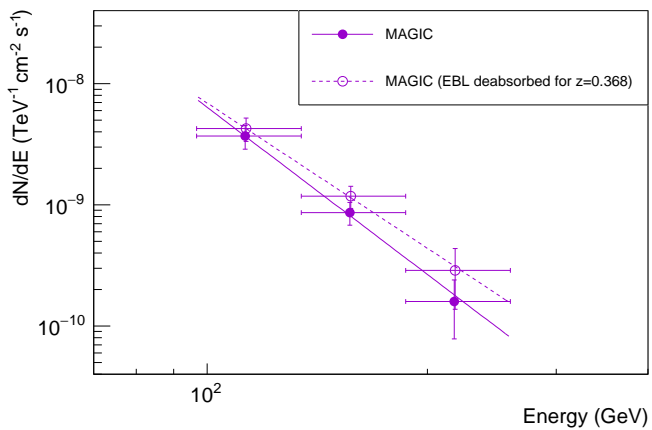}}
        \caption{Spectrum for the VHE MAGIC detection. MAGIC data are for flare night only (2015 February 14, MJD 57067.14). Violet filled circles are for the unfolded observed points, while open circles are de-absorbed for EBL absorption \citep[EBL model by][]{dominguez11}. The solid line is the fit for the observed points and the dashed line is the fit for the de-absorbed ones, with details in the text.}
         \label{fig:magic_spec}
\end{figure}

The \gr emission from sources at high redshift is absorbed via photon-photon pair production on photons from the extragalactic background light \citep[EBL, see e.g.][]{dominguez11,finke2010}. \s4 redshift is assumed to be $z=0.368$.
The spectral shape of the intrinsic emission, i.e. after the correction for the EBL absorption, can be fitted with a simple power law:
   
   \begin{equation}
   \frac{dN}{dE} = N_0 \left(\frac{E}{E_0}\right)^{-\Gamma}  
   \end{equation}
with normalization $N_0 = \left(13.8\pm2.1^{\rm stat}\pm 1.5^{\rm sys}\right)\times 10^{-10} \rm{\ TeV}^{-1} {\rm cm}^{-2} {\rm s}^{-1}$ at $E_0=0.15$ TeV and spectral index $\Gamma=3.98\pm 0.67^{\rm stat}\pm 0.15^{\rm sys}$. The quoted systematic uncertainties are derived from the standard evaluation in MAGIC data presented by \citet{magicperformancepaper}. Note that the calculated systematic uncertainty on $N_0$ does not contain the uncertainty on the energy scale, that is about 15\%. The unfolded MAGIC spectrum is shown in Fig.~\ref{fig:magic_spec}. 
The unfolded observed spectrum, i.e. without correcting for the EBL absorption, can be described also by a simple power law with $N_0 = \left(9.9\pm1.5^{\rm stat}\pm 1.1^{\rm sys}\right)\times 10^{-10} \rm{\ TeV}^{-1} {\rm cm}^{-2} {\rm s}^{-1}$ at $E_0=0.15$ TeV and spectral index $\Gamma=4.58\pm 0.66^{\rm stat}\pm 0.15^{\rm sys}$.

\section{The Multiwavelength coverage}\label{sec:mwlcoverage}
All the data presented in this section are collected to produce the light curves and SED, whose interpretation is later presented in Section \ref{sec:lightcurveandsed}.
   
\begin{figure*}[p]
   \resizebox{\hsize}{!}
    {\includegraphics[]{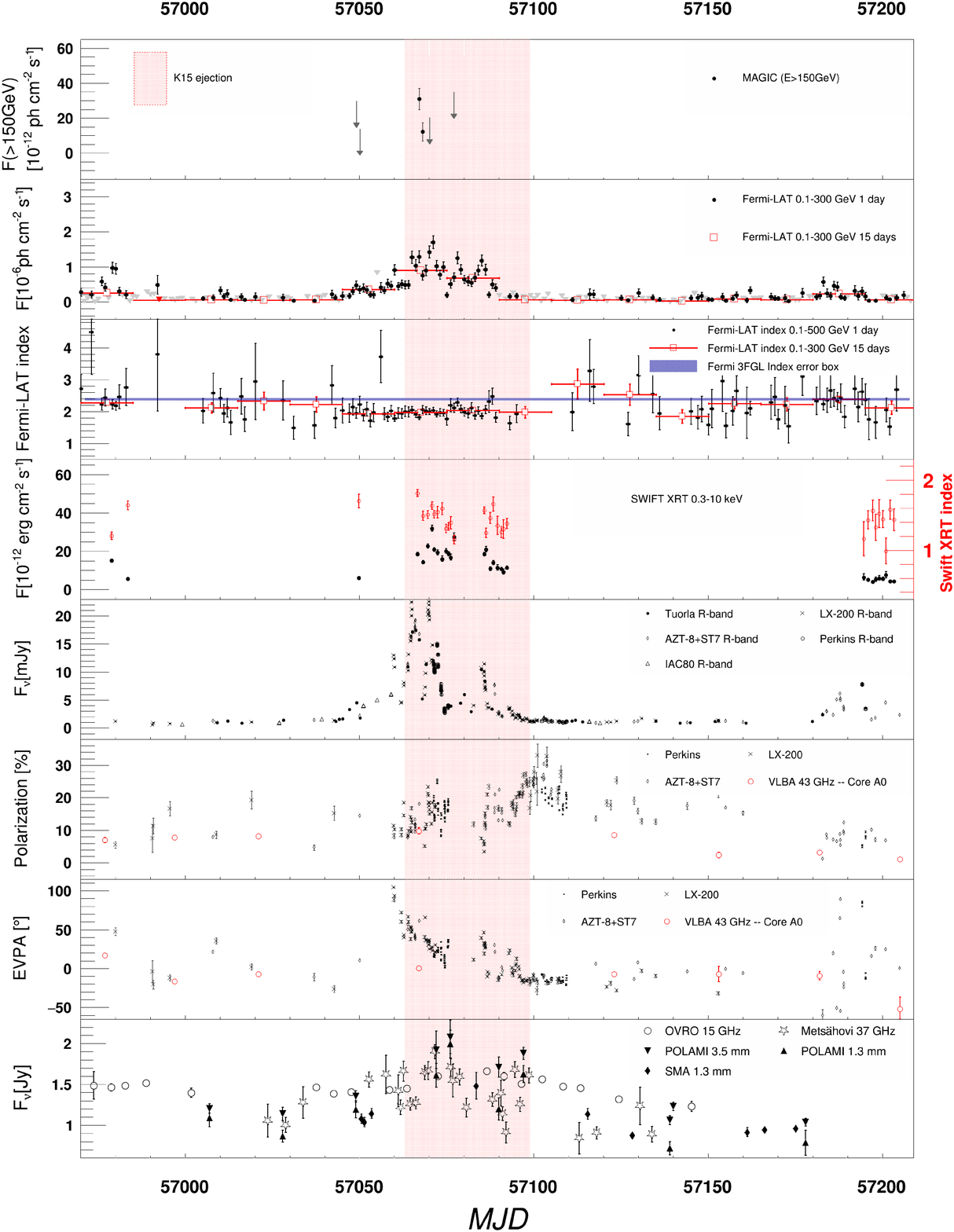}}
        \caption{MWL light curves and polarization evolution of \s4 ranging from MJD 56970 (2014 November 9) to MJD 57200 (2015 June 27). The energy range of each panel and the corresponding instrument can be found in the legend. Please refer to the text for details on the data taking and reduction for each instrument. }
         \label{fig:mwl_lc}
   \end{figure*}

\subsection{Fermi-LAT}
The LAT on board the \textit{Fermi} satellite scans the entire sky every 3 hours. From the data of the first 4 years of operation, \s4 was detected with an average significance of $27.2\sigma$ in the energy range from 100 MeV to 300 GeV as reported in the 3FGL catalog \citep{3FGL}. A dedicated analysis from MJD 56952 (2014 October 22) to MJD 57208 (2015 July 05) is presented in this work. We selected Pass 8 source class events within a 10$^\circ$ circular region centered on the position of \s4, in the energy range 0.1-500 GeV. The spectral analysis was performed through an unbinned likelihood fit, using the ScienceTools software package version v11-05-00 along with the instrument response functions P8R2\_SOURCE\_V6. The model of the likelihood fit includes a Galactic diffuse emission model and an isotropic component\footnote{Model available at \url{https://fermi.gsfc.nasa.gov/ssc/data/access/lat/BackgroundModels.html}.}. In addition, we included the sources in the 3FGL catalog within a 20$^\circ$ circular region centered on \s4. The spectral indexes and fluxes of the 3FGL sources located within a region of 10$^\circ$ from \s4 were left free to vary, while the sources in the region from 10$^\circ$ to 20$^\circ$ were fixed to their catalog values. The results were obtained from two iterations of maximum-likelihood analysis, after the sources with a test statistics \citep[][]{egretlikelihood} TS$<$10 were removed. The strongest source located beyond 10$^\circ$ from \s4 is at an angular distance of 10.8$^\circ$. This source has a variability index of 42.4 in the 3FGL catalog, that allows us to treat it as a non-variable source and thus to fix its spectral index and flux to the values reported in the 3FGL catalog.

The light curve was calculated in day timescale bins, modeling the source with a single power-law spectrum (as it is also described in the 3FGL). Both the flux and spectral index of \s4 were left free during the likelihood fits, while the rest of the point sources were fixed and only the diffuse Galactic and isotropic models were allowed to vary. In case of TS<4, an upper limit on the flux was calculated fixing the spectral index to 2.38 as given in the 3FGL catalog. The results are shown in Fig. \ref{fig:mwl_lc}. The figure shows also the light curve calculated in 15-days bin as comparison. The light curve was obtained with the same procedure described above for the 1-day binning. During the HE flare in November 2014 \citep[MJD 56976, ATel \#6709;][]{atel6709} the LAT spectral index is compatible with its 3FGL value of $2.38 \pm 0.04$, averaged from 4 years of data. Moreover, the visibility of the source by MAGIC was at an unfavorable zenith angle of 60$^\circ$ (implying a high energy threshold). Therefore, no ToO observation was activated with MAGIC for this flare. MAGIC observations were activated later on during the strong flare on February 2015 when the LAT detected a hardening of the spectrum as shown previously by Tanaka et al. (2016) where the LAT analysis using Pass 7 reprocessed data is presented.

The spectral analysis for the MWL SED corresponds to 1-day integration centered in the MAGIC observation (MJD 57067.14, 2015 February 14). From a first likelihood fit we found the best spectral fit was a power-law spectral index of $1.87 \pm 0.09$ (significantly harder than its average 3FGL value) and was fixed in the model for the spectral points calculation. Moreover, all the sources included in the model except the diffuse Galactic and isotropic models were also fixed. The source was detected during this period with a TS of 379.7. A curved spectral model is not significantly favored in this day (TS for a log parabola fit is TS$_\textrm{LP}$=380.10 to be compared with a simple power law fit with TS$_\textrm{PWL}$=379.74).

\subsection{\textit{Swift} dataset}

The 22 multi epochs event-list obtained by the X-ray Telescope \citep[XRT,][]{swiftXRT} on board the \textit{Swift} satellite in the period of 2014 November 17 (MJD 56978.96395) to 2015 March 11 (MJD 57092.26632) with a total exposure time of $\sim$11.12 hours were processed using the procedure described by \citet{vandad2017}. All these observations had been performed in photon counting (PC) mode, with an average integration time of 1.8 ks each. The equivalent Galactic hydrogen column density is fixed to the value of $n_H = 5.17 \times 10^{20} \textrm{[cm}^{-2} \textrm{]}$ \citep{LABHI}.

The average integral photon X-ray flux (0.3-10 keV) in this period is $1.64 \times 10^{-11}$ $\rm erg/cm^2/s$. The X-ray flux is peaking at MJD 57070.76523 with $F_{(0.3-10 \mathrm{keV})} = 3.18 \times 10^{-11}$ $\rm erg cm^{-2} s^{-1}$ which is a factor of about 2 higher than the average flux of the analyzed period. The average flux outside the flare period (2006-2015) is $F_{(0.3-10 keV)} = 4.3 \times 10^{-12}$ $\rm erg cm^{-2} s^{-1}$, that we derived from a sample of XRT data comprising 25 X-ray exposures in the XRT database, not including the 22 multi epochs event-list described above. This indicates that the source was clearly in its X-ray high state during the VHE $\gamma$-ray detection. The X-ray spectral index during the analyzed period varies between $1.15\pm0.06 \le \Gamma_X \le 1.82\pm0.1$. It is notable that the softest spectral index was obtained a night prior to the VHE $\gamma$-ray flare while the spectra starts to harden after 2015 February 14 and reach its historical hardest spectra 10 days after the VHE $\gamma$-ray flare. The X-ray spectra on the night before and after the VHE $\gamma$-ray flare can be well described with a power-law with spectral index of $\Gamma_{\rm X, Feb. 13} = 1.82 \pm 0.05$ ($ \chi^2/$d.o.f.=1.024/41) and $\Gamma_{\rm X, Feb. 15} = 1.49 \pm 0.07$ (1.025/24 $ \chi^2/$d.o.f.) respectively. 
The full dataset analysis is given in Appendix \ref{app:swiftxrt}.

The \textit{Swift} satellite hosts an additional instrument, the  Ultraviolet/Optical Telescope \citep[UVOT, ][]{uvotmainref}. The data taken during the period of interest for this work have already been presented by \citet{tanaka2016}. They follow the behavior of the optical light curve that we will present next. Therefore they are not reproduced again nor shown in Fig. \ref{fig:mwl_lc}. The UVOT bands are however important for the SED modeling presented in Section \ref{sec:lightcurveandsed} and will therefore be included there for MJD 57067 (2015 February 14, day of the VHE detection). The dataset presented by \citet{tanaka2016} suffers from an incorrect exposure calculation by a factor of 2, related to the deadtime correction, and thus a lower reconstructed flux. We therefore have performed a re-analysis here for the two exposures taken with UVOT on MJD 57066.76. Data reduction has been done on all the available filters ($v,\ b,\ u,\ w1,\ m2,\ w2$), following the standard UVOT data analysis prescriptions\footnote{\url{https://swift.gsfc.nasa.gov/analysis/}}. We present both exposures separately, due to the high variability in this night (e.g for the V-band there is a variation of $\sim$0.3 magnitudes in $\sim$1.5 hours).

\subsection{The optical domain}\label{subsec:optical}

Optical data were collected with: 35cm KVA telescope (La Palma Island, Spain) used in the Tuorla monitoring program; 1.8~m Perkins telescope of Lowell Observatory (Flagstaff, Arizona); 70 cm telescope AZT-8 at the Crimean Astrophysical Observatory (Nauchny, Russia); 40 cm telescope LX-200 of St. Petersburg State University (St. Petersburg, Russia); IAC80/Camelot at the Teide Observatory (Tenerife, Spain). The data analysis from KVA was
performed with the semi-automatic pipeline using the standard analysis procedures (Nilsson et al. in prep). The differential photometry was performed using the comparison star magnitudes from \citet{villata}. For the Perkins telescope see \citet{jorstad2010} and references therein. The details of observations and data reductions with AZT-8 and LX-200 are given by \citet{larionov2008}. IAC80/Camelot data were automatically processed by the pipeline Redcam and calibrated astrometrically using XParallax, both available at the telescope. Instrumental magnitudes for IAC80/Camelot data were extracted using Sextractor \citep{bertin96} and calibration of the source magnitude was obtained with respect to the reference stars provided by \citet{raiteri}.

All the above mentioned telescopes provide R-band photometry. We have applied the calibration of \citet{mead1990} for all optical measurements to transform magnitudes into flux densities, and dereddened the flux according to the absorption by \citet{schlafly}. The host galaxy is not detected for this object.

From the Perkins, AZT-8+ST7 and LX-200 telescopes we collect also polarization information. In Fig.~\ref{fig:mwl_lc} we show the optical photometry data and time evolution of the fractional linear polarization and the electrical vector position angle (EVPA) in R-band. The EVPA measurements have been arranged such to minimize the impact of the $\pm180^\circ$ ambiguity, i.e. adding or subtracting $180^\circ$ whenever two subsequent measurements differ by more than $90^\circ$. 

In the same timeframe of the VHE detection and the optical flare, a substantial change in the optical EVPA can be identified (see Fig. \ref{fig:mwl_lc}). The EVPA rotation starts just before the optical and VHE flare and reaches a total change of roughly 100$^\circ$. 
The optical flare in February 2015 is a factor of about 3 larger in flux than the 2011 flare \citep[see][]{morozova2014}, that was already exceptional and concurrent with a series of \gr flares evident in \fermi data. During the most extreme flare in 2011, the EVPA rotated by about 300$^\circ$.

\begin{figure*}[!htb]
   \resizebox{\hsize}{!}
            {\includegraphics[]{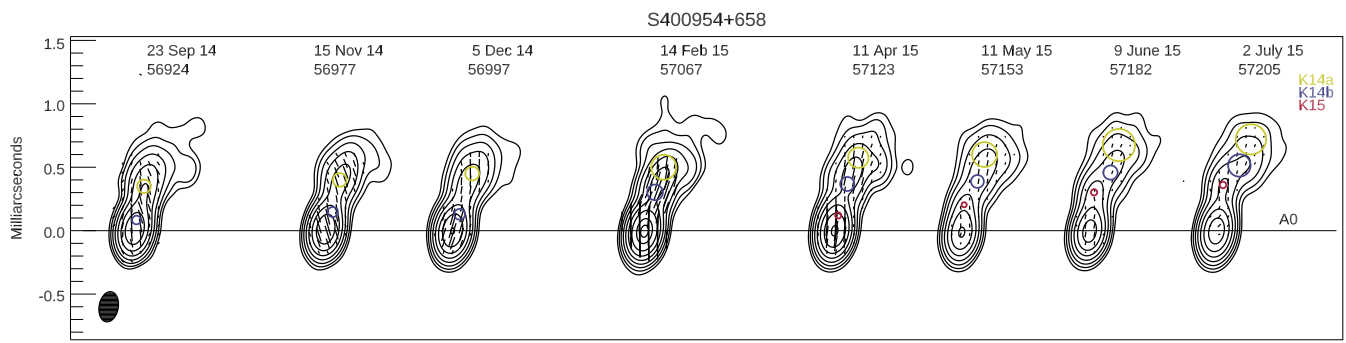}}
      \caption{ A sequence of total (contours) and polarized (segments) intensity images of S4 0954+658 at 43 GHz, convolved with a beam of 0.24$\times$0.15 mas$^2$ at PA=-10$^\circ$. The global total intensity peak is 1606 mJy/ beam and the global polarized intensity peak is 104 mJy/ beam; black line segments within each image show the direction of polarization and their length is proportional to the polarized intensity. The black horizontal line indicates the position of the core, A0, and grey, blue, and red circles show the locations of knots K14a, K14b, and K15, respectively. The size of the circles is proportional to the estimated average size in each epoch. The detailed characteristics of the knots can be found in Table \ref{table:knots} and in Table \ref{table:knots_LC}. }
         \label{fig:43ghzimages}
   \end{figure*}

\begin{figure}[!htb]
   \resizebox{\hsize}{!}
	  {\includegraphics[]{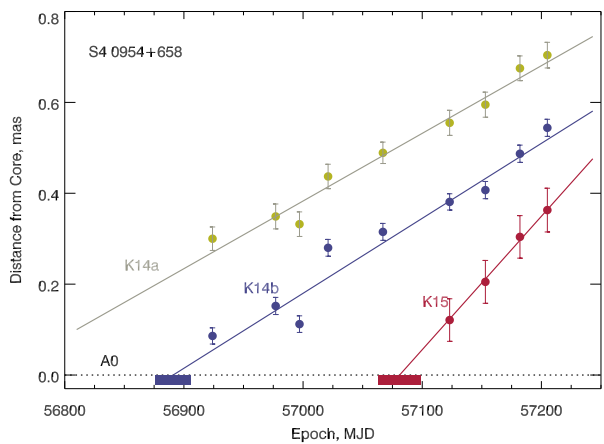}}
    \caption{Apparent distance from the radio core A0 of the new emerging knots, K14a,b and K15, as a function of time. The images from which the apparent distances are calculated can be found in Fig. \ref{fig:43ghzimages}. It can be noted that the K15 knot presents the highest apparent speed.}
         \label{fig:43ghzknots_characteristics}
    \end{figure}
   
\begin{table*}[!htb]
\caption{Characteristics of the new radio knots observed from the jet of \s4. The evolution of parameters with the monitoring snapshots can be found in Appendix \ref{app:vlbadata}. }             
\label{table:knots}      
\centering                          
\begin{tabular}{c c c c c c c c}        
\hline\hline                 
Knot & Average Flux & Maximum Flux & Average PA & Average Size & Proper motion & Apparent Speed & Time of Ejection \\
 & mJy & mJy & deg ($^\circ$) & (FWHM) mas & mas/yr & c & MJD \\
\hline                        
K14a & $120\pm7$ & $286\pm10$ & $-17.6\pm2.4$ & $0.15\pm0.07$ & $0.55\pm0.04$ & $12.49\pm0.91$ & $56708\pm26$ \\
K14b & $76\pm25$ & $118\pm6$ & $-16.2\pm2.6$ & $0.07\pm0.06$ & $0.59\pm0.04$ & $13.47\pm0.86$ & $56891\pm15$ \\
K15 &$109\pm14$ & $121\pm5$ & $-5.9\pm1.9$ & $0.05\pm0.01$ & $1.11\pm0.08$ & $25.27\pm1.20$ &$57081\pm18$ \\
\hline                                   
\end{tabular}
\end{table*}
\subsection{The radio and millimeter ranges}
S4 0954+65 was monitored at 3.5 mm (86 GHz) and 1.3 mm (229 GHz) wavelengths from the IRAM 30 m Millimeter Radiotelescope under the POLAMI (Polarimetric Monitoring of AGN at Millimeter Wavelengths)\footnote{\url{https://polami.iaa.es}} program. The program monitors the four Stokes parameters of a sample of the brightest ~40 northern blazars with a cadence better than a month 
\citep[see][]{agudo2018a,agudo2018b,thum2018}. Results from the observations are presented in Fig. \ref{fig:mwl_lc}. The data reduction, calibration, and flagging procedures were described in detail by Agudo et al. 2017a, submitted \citep[see also][]{agudo2010,agudo2014}. Fig. \ref{fig:mwl_lc} includes also the 1.3mm flux density data that were obtained at the Submillimeter Array (SMA) located in Hawaii. \s4 is included in an ongoing monitoring program at the SMA to determine the fluxes of compact extragalactic radio sources that can be used as calibrators at mm wavelengths \citep{gurwell2007}.  Observations of available potential calibrators are from time to time observed for 3 to 5 minutes, and the measured source signal strength calibrated against known standards, typically solar system objects (Titan, Uranus, Neptune, or Callisto). Data from this program are updated regularly and are available at the SMA website\footnote{\url{http://sma1.sma.hawaii.edu/callist/callist.html}}. The largest flux in the considered period is at MJD 57072-57076, showing an increase of the flux at 1 mm and 3 mm wavelengths. 
It is to be noted however the lack of exactly simultaneous data to the MAGIC peak detection (MJD 57067).

\s4 is monitored monthly by the Boston University (BU) group with the Very Long Baseline Array (VLBA) at 43 GHz within a sample of bright \gr blazars through the VLBA-BU-BLAZAR program\footnote{\url{http://www.bu.edu/blazars/VLBAproject.html}}. The VLBA data are calibrated and imaged in the same manner as discussed by \citet{jorstad2005,jorstad2017}. 
The VLBA imaging monitoring program allows us to study the kinematics of the inner jet at pc scale. The inner jet has been monitored also for months after the VHE flare (see Fig. \ref{fig:43ghzimages}). In addition to the stable core at mm wavelengths (dubbed A0, see Fig. \ref{fig:43ghzimages}) it was possible to identify the emergence of three new knots whose characteristics are tabulated in Table \ref{table:knots}. The nomenclature of the knots follows in sequential order from the beginning of the VLBA monitoring program. Previous knots characteristics can be found in \citet{morozova2014}.

Of particular interest is knot K15, which is very compact, with a FWHM average size of $0.05\pm0.01$ mas and presents the largest apparent speed of ($25.27\pm1.20$)c, cf Fig. \ref{fig:43ghzknots_characteristics}. The zero-epoch separation of this knot is consistent with the VHE flare considering its 18-day uncertainty. The intensity of the core is increasing in the epoch of MJD 57067 observation, but no significant change in the core polarization can be appreciated. The detailed information on the time evolution of the radio knot can be found in Table \ref{table:knots_LC}, while the polatization evolution details are shown in Table \ref{table:CorePol_LC}. During November 2014, while the source was high in the HE band as observed by \fermi but without optical enhancement, no new knot appears. The zero epoch-separation from the core of knots K14a and K14b are not coincident with the high state in \fermi data of November 2014, but happen months before. We analyzed \fermi data in the period included within the error band for K14a and K14b zero epoch-separation and found no particular enhancements. 

To be noted is also the position angle of K15 with respect to the core, (PA=$-5.9^\circ\pm1.9^\circ$). This is different than the values reconstructed from previous knots, ranging from roughly PA=$-15^\circ$ to PA=$-25^\circ$ in \citet{morozova2014}, that are in turn consistent with the values for K14a/b. The mean jet direction is at PA$\simeq -20^\circ$. A difference in PA and in apparent speed could be simply related to a small difference in the angle to the observer. However, the highest apparent speed can be used to estimate the Doppler factor, considering the upper limit to largest possible viewing angle $\theta_\textrm{obs}<\textrm{arcsin}(1/\beta_\textrm{app})$ and ultimately leading to $\delta_\textrm{app}\sim\beta_\textrm{app}$. Applying this to the above mentioned knots (averaging the apparent speed to $\beta_\textrm{app}\sim13$c for K14a/b): $\theta_\textrm{obs,K15}<2.3^\circ$ and $\delta_\textrm{app,K15}\sim25$; $\theta_\textrm{obs,K14}<4.4^\circ$ and $\delta_\textrm{app,K14}\sim13$.

The 37 GHz observations were made with the 13.7 m diameter telescope at Aalto University Mets\"ahovi Radio Observatory. A detailed description of the data reduction and analysis is given by \citet{metsa}. The error estimate in the flux density includes contributions from the measurement RMS and the uncertainty of the absolute calibration. The S4 0954+65 observations were done as part of the regular monitoring program and the GASP-WEBT campaign. There are no strictly simultaneous 37GHz data to the MAGIC detection, however an increase in flux can be seen when comparing observation taken one day before (2015 February 13, MJD 57066.15, $F_\nu=1.27\pm0.07$ Jy) and one day after the MAGIC detection (2015 February 15, MJD 57068.15, $F_\nu=1.65\pm0.09$ Jy).

The OVRO 40 m uses off-axis dual-beam optics and a cryogenic pseudo-correlation receiver with a
15.0 GHz center frequency and 3 GHz bandwidth. 
Calibration is achieved using a temperature-stable diode noise source to remove receiver gain
drifts and the flux density scale is derived from observations of
3C~286 assuming the \citet{1977A&A....61...99B} value of 3.44 Jy at
15.0 GHz. The systematic uncertainty of about 5\% in the flux density
scale is not included in the error bars.  Complete details of the
reduction and calibration procedure are found in \citet{ovro}.
The long-term monitoring program at OVRO (Owens Valley Radio Observatory) monitors the variability of this source at 15GHz over a longer time than what shown here. While it is obvious  
that the source was variable also during February 2015, it is not an exceptionally bright flux state of the source in the radio band. From a decade long monitoring, the source shows brighter levels (highest at $F_\textrm{15 GHz} =2.53$ Jy) and fainter levels (lowest at $F_\textrm{15 GHz} =0.85$ Jy).

Both 15 GHz and 37 GHz data seem to be in agreement with the behavior seen from mm wavelength data. Again note the lack of strictly simultaneous data to the MAGIC peak detection (MJD 57067). emission.

\section{Discussion}\label{sec:lightcurveandsed}
The coverage of flaring states at VHE is helpful to understand the jet dynamics. We present a discussion of the SED for the day of the flare (2015 February 14). We do not attempt a SED modeling for other days, for which the MAGIC data would provide only non-constraining upper limits to emission at VHE. 
The day of the VHE detection is instead put in context with a longer time span behavior in the MWL dataset. However the VHE sampling of the state is too scarce to attempt a numerical correlation study of the light curves.
\subsection{Light curve phenomenology}
The MWL light curves of the source for all the instruments involved in the present work are reported in Fig. \ref{fig:mwl_lc}, and cover a time range of 7 months, from MJD 56970 (2014 November 19) to MJD 57200 (2015 June 27). The panels of Fig. \ref{fig:mwl_lc}, in order of decreasing energy, show in the top panel the MAGIC detection at VHE, while the radio data collected by OVRO, POLAMI and the other instruments in the radio band are shown in the bottom one. The red region indicates the time window where the knot K15 was ejected in the VLBA analysis, as reported in Table \ref{table:knots}: a time range of 36 days centered in MJD 57081 (2015 February 28). The VHE detection and the enhanced activity in the other bands are found inside the K15 ejection time window, making this event important for the understanding of the whole scenario. The spectral index at HE as inferred from the \textit{Fermi}-LAT data is harder than the average spectral index of $\Gamma=2.38\pm0.04$ from the 3FGL catalog dataspan. \
In the presented timeframe, the X-ray emission peaks around the observation on MJD 57070.76434 (2015 February 17), with a delay with respect to the detection in VHE. The $\sim$ 3 hours of observations in VHE during the same night did not lead to a detection (see Table \ref{table:magicdata}). However during the period of enhanced MWL activity, there is a clear hardening of the X-ray spectrum. Hardening at both X-ray and \gr energies points toward the emergence of a new component in the non-thermal spectrum. \par
\begin{figure}[!htbp]
   \resizebox{\hsize}{!}
    {\includegraphics[]{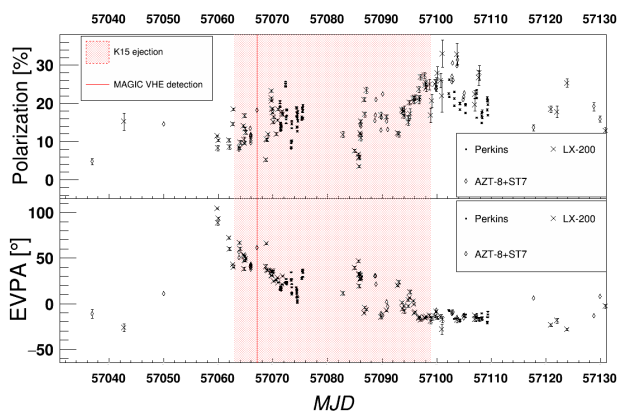}}
        \caption{Light curves for R-band polarimetry of \s4. Please refer to the text for details on the data taking and reduction for each instrument. }
         \label{fig:optical_lc}
   \end{figure}

The optical band is very bright during the VHE detection, reaching peaks of more than 20 mJy of flux density when the average behavior of the source is found around a few mJy (see the optical monitoring from Tuorla observatory).  
The optical emission is polarized by a fraction of $\gtrsim10\%$  and the polarization angle rotates by $\sim100^\circ$ during the flare: \citet{blinov2015} have shown that from a systematic monitoring (Robopol monitoring) of both \gr loud and \gr quiet sources, only the former class of object displays polarization angle rotation similar to the one seen here for \s4. \citet{blinov2015} studied the change of EVPA as a function of time for smooth changes of $>90^\circ$. Requesting the same smoothness requirements, no smooth rotation of $>90^\circ$ can be identified in the dataset presented here, see Fig.~\ref{fig:optical_lc}. A non smooth variation of $\Delta_\textrm{EVPA}\simeq105^\circ$ can however be identified between MJD 57060 and MJD 57075. This variation would imply a change of the EVPA curve slope of $\Delta_\textrm{EVPA}/\Delta_t=7 \textrm{deg/day}$,  compatible with the bulk of the variations studied by \citet{blinov2015}.
The rotations of the polarization angle are often physically linked to high flaring states of the objects in the \gr band. While individual occurrences of \gr flares and rotations cannot be firmly linked to each other, there is a low probability that all the occurrences are due to chance coincidence \citep[from MonteCarlo simulations in][]{blinov2015}. This hypothesis is still confirmed from 3 years of Robopol monitoring data in \citet{blinov2018}. \citet{kiehlmann2017} also study whether a simple stochastic variation can account for the observed rotations in the Robopol monitoring. While their model is failing to recover all the observational characteristics in the monitoring, it also highlights a larger discrepancy from the expectations of stochastic model with respect to the occurrence of large variations of EVPA ($>90^\circ$), however not significant. 
Smooth variations seem also to be more firmly linked to deterministic processes and not to a random walk effect \citep{kiehlmann2016}.
Robopol monitoring data are also used in \citet{angelakis2016}, to study the difference in the amount of polarization seen on average in \gr loud and \gr quiet sources. The median fraction variability of the \s4 dataset presented here is $16.4\%$. This value can be compared with the average $10\%$ for the \gr loud subset of the Robopol monitoring and a value of $17.1\%$ for \s4 computed for the observations on year 2013 and 2014. According to the interpretation by \citet{angelakis2016}, a higher fractional polarization is also expected in LSP/ISP blazars, due to the fact that in such sources the optical synchrotron emission relates to the peak synchrotron emission. Therefore, the particles associated with this emission are the most energetic, with faster cooling and thus probing a small volume of the emission region near the acceleration region, where it is expected to have a stronger ordered (helical) magnetic field, leading to higher polarization fraction.

Images at 43 GHz show the emergence of new knots. In \citet{morozova2014}, a series of optical flares of \s4 in 2011 are studied, and the emission of knots  is found correlated to the simultaneous flaring of the optical and HE bands. The maximum flux in the 2011 state is a factor of 3 lower in optical than the state presented here. The polarization fraction in this 2011 flare was similar to that seen in the present work. In \citet{morozova2014} the chance coincidence of high optical state and knot emission has very low probability. \par
The phenomenology of the 2015 flare described here agrees very well with the model put forward by \citet{marschernature} and applied to the \s4 dataset of \citet{morozova2014}. In that model, the flare is due to a newly appearing knot accelerating at the base of the jet and propagating through an helical flow streamline. The helical streamline can be expected due to the anchoring of the accelerating flow to the rotating base of the accretion disk or black hole magnetosphere, depending on modeling. The magnetic field topology in the jet is also helical and ordered. Geometrical effects and the propagation through the helical magnetic field account for the rotation of the EVPA. 

In \citet{zhang2014}, a model is proposed where the EVPA rotation is also related to the propagation through an helical magnetic field, but the streamline of propagation is not necessarily helical itself. In this model the magnitude of the swing can depend on the assumptions on the settings for the flare, specifically the magnetic field strength and orientation, the acceleration efficiency and the  continuous injection of freshly accelerated particles.

The model described in \citet{marschernature} allows the emission at radio wavelengths in a flaring state which is not simultaneous with the VHE flare. In this scenario the radio activity could be delayed several days, even months, with respect to the VHE detection. This is expected if synchrotron self absorption is involved, and hence the emission region is located closer to the central engine than the radio core (A0 in Fig. \ref{fig:43ghzimages}). The peak of radio emission is expected to be lagging behind and appear when the disturbance has propagated further down the jet, where the absorption is not an issue. The X-ray emission peak, then, could also be delayed with respect to the optical outburst. As the X-ray emission is probably due to IC of an external soft photon field by electrons in the jet (see above), the X-ray variability traces both the accelerated particle distribution and a change in the soft photon field. This retraces similar interpretation drawn for flares of other sources where the dataset was however richer and more detailed \citep{marschernature,marscherpks1510,magic15102012,magic15102017}. 

\subsection{Emission model for the flare SED}
\begin{figure*}[!htb]

    {\includegraphics[width=\textwidth]{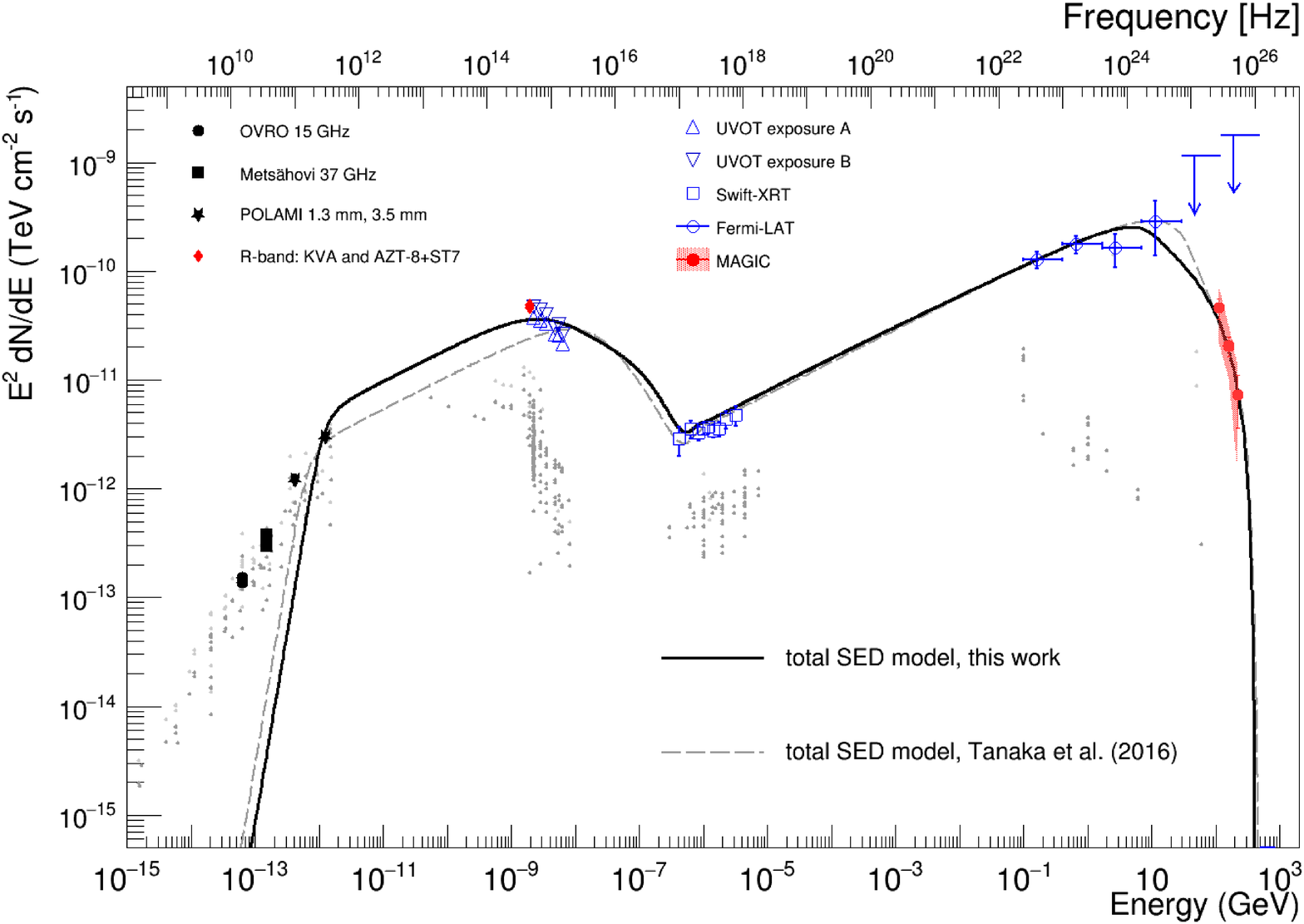}}

        \caption{Spectral energy distribution for the VHE MAGIC detection. Red symbols are strictly simultaneous to the VHE detection, blue symbols are for data taken during the same day and black symbols are for the closest observations. MAGIC spectral data (red circles) are for flare night only (2015 February 14, MJD 57067.14). Red filled circles are for the unfolded observed data points. The red shaded band shows the region of additional systematic uncertainty. \fermi data are the PASS8 data for 2015 February 14 (1-day integration centered on the MAGIC observation, blue squares). \textit{Swift}-XRT data are for 2015 February 13 (MJD 57066.70992, blue squares). \textit{Swift}-UVOT data are given for the two separate exposure taken on 2015 February 13 (MJD 57066.76, blue triangles and dark blue triangles). R-band data are for 14th Feb (Tuorla, MJD 57067.16375 and AZT-8+ST7 MJD 57067.1, red diamonds). POLAMI data are for the 18th February (MJD 57071.5, black stars at 100 GHz and 300 GHz). OVRO data are for 2015 February 10 and 19 (black circles at 15 GHz). Mets\"{a}hovi for 2015 February 13 and 15 (MJD 57066.15, MJD 57068.68, black squares at 37 GHz). The gray data are for NED (light) and SSDC (dark) SED historical data points. The model from \citet[][gray dashed curve]{tanaka2016} as well as the model presented here (black solide curve) include an emission component from synchrotron plus inverse Compton on a dusty torus (see text for details and Table \ref{table:SEDmodel} for the values of the physical parameters). The effect of the EBL attenuation is included in the modeling using the model by \citet{finke2010} and a redshift of z=0.368.}
         \label{fig:mwl_sed}
\end{figure*}

The SED of blazars are dominated by their non-thermal emission and can usually be described by two broad components. The low energy non-thermal emission is explained as synchrotron emission, while the high energy emission is most commonly modeled through inverse Compton (IC) emission, where soft photons are upscattered to $\gamma$-ray energies by electrons within the jet emitting region. The origin of the soft photon field itself can vary for different blazar subclasses. In particular, for most of the classical BL Lac objects, the VHE emission can be reasonably modeled through Synchrotron self-Compton emission \citep[SSC, see e.g.][]{rees67,maraschi92}. Instead, for the case of FSRQs, the modeling of the emission usually requires the inclusion of external soft photon fields from e.g. the infrared dusty torus or the optical-ultraviolet emission from the Broad Line Region (BLR) for the IC process \citep[see e.g.][]{tavecchiogamma}.

A broadband SED is compiled for 2015 February 14 (MJD 57067). We collect, from the MWL sample described in Section \ref{sec:mwlcoverage}, the data closest in time to the MAGIC observation. \fermi data points are obtained from a 1-day integration centered on the MAGIC observation. The specific dates of other wavelength observations are given in the caption of Fig. \ref{fig:mwl_sed}.

\citet{tanaka2016} model the SED of \s4 during a similar integration time as the 2015 flare studied in this work. The data shown in Fig. \ref{fig:mwl_sed} include, in addition to what is shown by \citet{tanaka2016}, the VHE data from the MAGIC observation, the AZT-8+ST7 and POLAMI data. Moreover, the \fermi data are reanalyzed as described in Section \ref{sec:mwlcoverage} to be centered at the MAGIC observation time and benefit from the latest \fermi PASS8. The \textit{Swift}-XRT and \textit{Swift}-UVOT data are also reanalyzed for this work. 

\citet{tanaka2016} report that a SSC modeling of the data is challenging, requiring very low magnetic field ($\textrm{B}\sim1\mu$G in contrast to the $\textrm{B}\sim1G$ expected in blazar jet components). Alternatively, an External Compton (EC) modeling was able to reproduce the data. In their model, the soft photon field for the EC {model} was the dusty torus from the source. In Fig. \ref{fig:mwl_sed}, we plot the model from \citet{tanaka2016}. This model reproduces the \fermi and MAGIC data, although their paper did not include any MAGIC data. However, the model fails to reproduce properly the optical observations. Such underestimation at optical frequencies in the model of \citet{tanaka2016} is driven by a misreconstruction of the UVOT fluxes, explained in Section \ref{sec:mwlcoverage}. With the reanalyzed UVOT dataset presented here, we use a new model, using the same code and most of the same assumptions as in \citet{tanaka2016}, including a redshift of $z=0.368$. The code is explained in detail in \citet{finke08} and \citet{dermer09}. Note that the presented SED model curves already include the effect of EBL absorption, i.e the intrinsic emission is absorbed according to the EBL model by \citet{finke2010}. The new EC model provides a good description of the MWL data and is shown in Fig. \ref{fig:mwl_sed}. The parameters of both models are reported in Table \ref{table:SEDmodel}. The break in the underlying electron population is similar to what expected by classical cooling, with the slope of the electron distribution before of the break ($s_1$) and after the break ($s_2$) differing by $s_2-s_1=1.2$. The use of VHE spectral information is crucial to model the falling part of the high energy peak of the blazars SEDs, which is crucial to constrain the most energetic electrons within the leptonic framework scenario (SSC and EC models).

\begin{table*}[!htbp]
\caption{SED model parameters}             
\label{table:SEDmodel}      
\centering                          
\begin{tabular}{c c c c}        
\hline\hline                 
Parameter & Symbol & Model A & Model B \\
          &        & \citet{tanaka2016} & this work \\
\hline
Redshift & 	$z$	& \multicolumn{2}{c}{0.368} \\
Bulk Lorentz Factor & $\Gamma$	& 30 & 35 \\
Doppler factor & $\delta_D$	& 30 &35 \\
Variability Timescale [s]& $t_v$       & $1.0\times10^5$ & $4\times10^4$\\
Comoving radius of blob [cm]& $R^{\prime}_b$ & 6.6$\times$10$^{16}$ & 3.0$\times$10$^{16}$\\
Magnetic Field [G]& $B$         & 0.6 & 0.4 \\
\hline
Low-Energy Electron Spectral Index & $s_1$   & 2.4 & 2.4 \\
High-Energy Electron Spectral Index  & $s_2$ & 4.5 & 3.6 \\
Minimum Electron Lorentz Factor & $\gamma^{\prime}_{\rm min}$  & $1.0$ & $1.0$ \\
Break Electron Lorentz Factor & $\gamma^{\prime}_{\rm brk}$ & $8.0\times10^3$ & $4.0\times10^3$  \\
Maximum Electron Lorentz Factor & $\gamma^{\prime}_{\rm max}$  & $2.0\times10^4$ & $4.0\times10^4$ \\
\hline
Black hole Mass [$M_\odot]$ & $M_{\rm BH}$ & \multicolumn{2}{c}{$3.4\times10^8$} \\
Disk luminosity [$\erg \ \mathrm{s}^{-1}$] & $L_{\rm disk}$ & \multicolumn{2}{c}{$3.0\times10^{43}$} \\
Inner disk radius [$R_g$] & $R_{\rm in}$ & \multicolumn{2}{c}{$6.0$}\\
Seed photon source energy density [$\erg\ \mathrm{cm}^{-3}$] & $u_{\rm seed}$ & $2.4\times10^{-4}$ & $4.4\times10^{-5}$\\
Seed photon source photon energy [$m_e c^2$ units] & $\epsilon_{\rm seed}$ & $7.5\times10^{-7}$ & $5\times10^{-7}$\\
Dust Torus luminosity [$\erg\ \mathrm{s}^{-1}$] & $L_{\rm dust}$ & $3.9\times10^{42}$ & $1.5\times10^{42}$ \\
Dust Torus radius [cm] & $R_{\rm dust}$ & $2.1\times10^{17}$ & $6.1\times10^{17}$ \\
Dust temperature [K] & $T_{\rm dust}$ & $1500$ & $1000$\\
\hline
Jet Power in Magnetic Field [$\erg\ \mathrm{s}^{-1}$] & $P_{j,B}$ & $1.0\times10^{46}$ & $1.4\times10^{45}$    \\
Jet Power in Electrons [$\erg\ \mathrm{s}^{-1}$] & $P_{j,e}$ & $1.1\times10^{45}$ & $6.6\times10^{45}$  \\
\hline                                   
\end{tabular}
\end{table*}

As mentioned in the introduction, the classification of a blazar can be aided by the study of its SED characteristics. According to the SED model presented above, the peak of the synchrotron emission is at $\nu_\textrm{syn}\sim8\times10^{14}$ Hz, making it an intermediate synchrotron peaked BL Lac object \citep{3LAC}\footnote{intermediate-synchrotron-peaked  blazar (ISP) are defined with rest-frame synchrotron peak frequencies of $10^{14}\textrm{Hz}<\nu_\textrm{syn}<10^{15}\textrm{Hz}$}. The Compton dominance, calculated comparing the luminosity at the peak of the synchrotron emission to that of the IC peak, is $L_\textrm{IC}/L_\textrm{syn}\sim 7$. Such Compton dominance value is at least 3.5 times the values obtained by \cite{finke_cd} for long-term blazar studies.

\section{Conclusions}

The census of extragalactic objects that present VHE emission is still limited. We present here the first detection at VHE of the blazar \s4 obtained through observations with the MAGIC Telescopes. The observations were conducted during an exceptional flare of the source in February 2015, originally identified in the optical band. We collected MWL simultaneous data to better characterize the state of the source.

The HE emission is also found in elevated state from the analysis of \fermi data, which reveal the hardest state of the HE emission to be concurrent with the detection at VHE. The X-ray emission peak is delayed by a few days with respect to the VHE detection and shows a trend of spectral hardening during the period presented here. The radio and mm wavelength emission reveal a moderate elevation of the flux, that is however not exceptional in the long term behavior of the source.

The source is classified in the literature as a BL Lac, but we have shown here that it presents similarities with the FSRQ class. Results from the monitoring of optical polarization and 43 GHz jet component analysis were compared to archival observation of \s4 and of statistical behaviour of other sources. Three main measurements were considered: the day of the VHE detection of \s4 is included in the error box for the zero epoch separation of knot K15; the optical polarization fraction is increasing in the same period; a rotation of optical EVPA of $\sim100^\circ$ can be identified, also in the same period, possibly related to the helical structure of the magnetic field in the acceleration region. We have discussed how these measurements point to a common behaviour with ISP/LSP sources. Both the best emission model (EC on dust torus) and the MWL light curve behavior show points of contact with other sources that are either clear FSRQ (like PKS 1510-089) or are transitional objects (like BL Lac itself). This is also supported from the moderate Compton dominance in the SED model and the fact that the synchtrotron peak show that the source can be classified as ISP source.

The work presented here reiterates the importance of VHE \grn and detailed MWL studies of blazars during different flux states to test their intrinsic characteristics and shed light on the physical processes taking place within their jets.


\begin{acknowledgements}
We would like to thank the Instituto de Astrof\'{\i}sica de Canarias for the excellent working conditions at the Observatorio del Roque de los Muchachos in La Palma. The financial support of the German BMBF and MPG, the Italian INFN and INAF, the Swiss National Fund SNF, the ERDF under the Spanish MINECO (FPA2015-69818-P, FPA2012-36668, FPA2015-68378-P, FPA2015-69210-C6-2-R, FPA2015-69210-C6-4-R, FPA2015-69210-C6-6-R, AYA2015-71042-P, AYA2016-76012-C3-1-P, ESP2015-71662-C2-2-P, CSD2009-00064), and the Japanese JSPS and MEXT is gratefully acknowledged. This work was also supported by the Spanish Centro de Excelencia ``Severo Ochoa'' SEV-2012-0234 and SEV-2015-0548, and Unidad de Excelencia ``Mar\'{\i}a de Maeztu'' MDM-2014-0369, by the Croatian Science Foundation (HrZZ) Project IP-2016-06-9782 and the University of Rijeka Project 13.12.1.3.02, by the DFG Collaborative Research Centers SFB823/C4 and SFB876/C3, the Polish National Research Centre grant UMO-2016/22/M/ST9/00382 and by the Brazilian MCTIC, CNPq and FAPERJ.
The \textit{Fermi} LAT Collaboration acknowledges generous ongoing support
from a number of agencies and institutes that have supported both the
development and the operation of the LAT as well as scientific data analysis.
These include the National Aeronautics and Space Administration and the
Department of Energy in the United States, the Commissariat \`a l'Energie Atomique
and the Centre National de la Recherche Scientifique / Institut National de Physique
Nucl\'eaire et de Physique des Particules in France, the Agenzia Spaziale Italiana
and the Istituto Nazionale di Fisica Nucleare in Italy, the Ministry of Education,
Culture, Sports, Science and Technology (MEXT), High Energy Accelerator Research
Organization (KEK) and Japan Aerospace Exploration Agency (JAXA) in Japan, and
the K.~A.~Wallenberg Foundation, the Swedish Research Council and the
Swedish National Space Board in Sweden.
This work performed in part under DOE
Contract DE-AC02-76SF00515
This research has made use of the NASA/IPAC Extragalactic Database (NED),
which is operated by the Jet Propulsion Laboratory, California Institute of Technology,
under contract with the National Aeronautics and Space Administration.
Part of this work is based on archival data, software or online services provided by the Space Science Data Center - ASI.
The OVRO 40-m monitoring program is supported in part by NASA grants NNX08AW31G, NNX11A043G and NNX14AQ89G, and NSF grants AST-0808050 and AST-1109911
The research at Boston University was supported by NASA Fermi Guest Investigator program grant 80NSSC17K0694 and US National Science Foundation grant AST-1615796. The VLBA is an instrument of the Long Baseline Observatory.
The Long Baseline Observatory is a facility of the National Science Foundation operated by Associated Universities, Inc.
This paper is partly based on observations carried out with the IRAM 30 m Telescope. IRAM is supported by INSU/CNRS (France), MPG (Germany) and IGN (Spain). IA acknowledges support by a Ram\'on y Cajal grant of the Ministerio de Econom\'ia, Industria y Competitividad (MINECO) of Spain. The research at the IAA--CSIC was supported in part by the MINECO through grants AYA2016--80889--P, AYA2013--40825--P, and AYA2010--14844, and by the regional government of Andaluc\'{i}a through grant P09--FQM--4784.
St. Petersburg University team acknowledges support from Russian Science
Foundation grant 17-12-01029. 
The Submillimeter Array is a joint project between the Smithsonian Astrophysical Observatory and the Academia Sinica Institute of Astronomy and Astrophysics and is funded by the Smithsonian Institution and the Academia Sinica.
\end{acknowledgements}

%
%

\begin{appendix}
\section{Additional information on MAGIC data reduction}\label{app:magicdata}
\begin{table*}[!h]
\caption{MAGIC data summary for the observation of \s4 from 27th January to 1st March 2015. Days of observations are listed along with the data qualification (see text for details), length of observation and significance of detection for different analysis cuts. For detections, also the integrated flux above 150 GeV is given. In the instances of non-detection, we provide a 95\% confidence level upper limit.}             
\label{table:magicdata}      
\centering                          
\begin{tabular}{c c c c c c}        
\hline\hline                 
MJD & Observation Time & Significance & Significance FR & F(>150GeV)\\    
    &       [h]      &  $ \sigma$ &  $ \sigma$ &  cm$^{-2}$ s$^{-1}$\\
obs    &         &  LE: hadr<0.28 &  FR: hadr<0.16 &  \\
 condition   &         &  size>60phe &  size>300phe &  \\
\hline                        
   57049.176 (1)& 0.33 & 0.64  &0.43&$<3.0\times 10^{-11}$\\      
   57050.164 (1)& 0.68 & -0.82 &0.19&$<1.4\times 10^{-11}$\\
   57067.139 (1)& 2.05 & 7.98  &0.20&$(3.1\pm0.6)\times 10^{-11}$\\
   " (1+3)        & 2.86 & 7.26  &-0.09&-\\
   " (3)          & 0.80 & 0.35  &-0.67&$<5.0\times 10^{-11}$\\
   57068.154 (1)& 2.53 & 3.19  &1.16&$(1.2\pm0.5)\times 10^{-11}$\\
   57069.099 (2)& 0.32 & -0.04  & 0.42&  $<5.0\times 10^{-11}$ \\
   57070.147 (1)& 2.91 & 2.12  &1.00&$<2.0\times 10^{-11}$\\ 
   57077.098 (1)& 0.97 & 2.41  &1.62&$<3.5\times 10^{-11}$\\
   57082.153 (4)& 0.94 & 2.79  &-0.50&$F(>250GeV)<2.1\times 10^{-11}$\\   	
\hline                                   
\end{tabular}
\end{table*}

The MAGIC telescopes are supported by an extensive weather monitor program. Atmospheric transmission at different heights within the MAGIC field of view is obtained with the use of a LIDAR \citep[for details on this see][]{lidarpaper}. For data quality selection we consider the transmission measured at a height of 9 km, with $T_{\rm 9km}=1$ representing a perfectly clear sky and $T_{\rm 9km}=0$ a complete opacity. MAGIC can carry out observations also during partial moonlight, with the drawback of having a higher energy threshold and larger systematic errors due to a higher contamination from the elevated night sky background (NSB), see \citet{magicmoonpaper}. The brightness of the NSB can be monitored from the average current in the camera (DC).
\s4 was observed in a zenith range ranging from 35$^\circ$ to 50$^\circ$, for a total of 12.5 hours of data, of which $\sim$1 hour was lost due to bad weather.
In the following we will refer to the different observation conditions of our data set as follows:
\begin{enumerate}
\item {\it good dark data}: data taken with dark sky (DC < $1.5 \mu$A) and good atmospheric condition ($T_{\rm 9km}>0.85$), used for detection and spectral reconstruction;
\item {\it dark data needing atmospheric correction}: data taken with dark sky (DC < $1.5 \mu$A) but under non optimal weather conditions ($0.55<T_{\rm 9km}<0.85$), used for detection and spectral reconstruction after atmospheric correction;
\item {\it good low moon data}: data taken with elevated NSB due to moonlight ($1.5 \mu$ < DC < $4 \mu$A) and good atmospheric condition ($T_{\rm 9km}>0.85$), used only for detection in this particular dataset;
\item{\it good moon data}: data taken during high NSB due to moonlight  (DC>$4 \mu$A) and good atmospheric condition ($T_{\rm 9km}>0.85$), used only for detection.

\end{enumerate}

The subsample of dataset selected with condition (1) (9.48 h of good quality data) has been analyzed with the standard MAGIC analysis chain \citep{standardMARS}.

The subsample of dataset selected with condition (2) (0.32 h of data) follows the same analysis chain until the estimation of the energy for the events and evaluation of the flux. For this last step, the estimated energy and the effective area are corrected taking into account the enhanced atmospheric absorption \citep[for validation of the procedure see ][]{lidarpaper}.

The subsample of dataset selected with condition (3) is applicable only at the day of 14th February, with the first VHE detection. The detection can be claimed from dark data alone (i.e. selected with condition 1), but an extra 0.81 h of data were taken under low moonlight. Those data are presented here for completeness, but are not used for spectral reconstruction so not to increase the systematic error and energy threshold.

The subsample of dataset selected with condition (4) (0.94 h of data) requires a special analysis that takes care of the effect of moonlight on data taking, reconstruction and analysis. Details of the procedure can be found in \citet{magicmoonpaper}. 

The detailed breakdown of significances and estimated VHE fluxes is given in Table  \ref{table:magicdata}. Numbers are presented for the so-called low energy (LE) and full range (FR) cuts. The LE cuts are optimized for an energy range of $E\gtrsim100$ GeV and are particularly appropriate for steep spectrum sources, while FR cuts are optimized for an energy range of $E\gtrsim250$ GeV. 
The cuts are applied on 2 parameters: the ''size`` parameter, integrated charge (in photoelectrons) in the cleaned shower image; the ''hadronness`` parameter, computed from the gamma-hadron separation Random Forest (RF), with a value ranging from 0 for the most gamma-like images to 1 for the most hadron-like images. Indeed the standard MAGIC analysis chain relies on RF techniques to discriminate among gamma and hadronic shower and to better reconstruct the event directions. Lookup tables are used for energy estimation. This is achieved starting from a parametrization of the shower images in the detector.
The significance of signal is then calculated with Eq. 17 from \citet{lima} and using 5 regions of equal size and distance to the center of camera as the signal region for background estimation. Fluxes are calculated above an energy threshold of 150 GeV, which corresponds to the peak of the differential energy distribution of the excess events as a function of estimated event energy. The high energy threshold is due to the high zenith angle of the observation. Please note that for data of condition 4, strong moon, we apply an additional minimum cut in the ''size`` parameter (''size`` $>$ 150 phe) of the reconstructed Cherenkov image as prescribed by the moonlight-adapted analysis. This increases the energy threshold to a value of $\sim$250 GeV.  In case of non-detection, we provide 95\% confidence level upper limits to the flux, calculated following \citet{trolke}, considering a systematic error on flux estimation of 30\% \citep{magicperformancepaper}.

\section{Additional VLBA derived parameters}\label{app:vlbadata}

The detailed information on the time evolution of the radio knot can be found in Table \ref{table:knots_LC}, while the polatization evolution details are shown in Table \ref{table:CorePol_LC}.
\begin{table*}[!htbp]
\caption{Time evolution of characteristics of the new radio knots observed from the jet of \s4. For each identified component and for each epoch (i.e. observation), we present: flux, position with respect to core AO, projected size and position angle. }             
\label{table:knots_LC}
\centering                          
\begin{tabular}{c c c c c c c c c}        
\hline
\hline
Epoch  &   MJD  &   Flux(Jy) & x &    y  &  R(mas) & PA(deg) &  Size(mas) & Knot\\
\hline
23 Sep 2014&&&&&&&&\\
2014.7288 & 56924& 0.558& 0.000&0.000&0.000&0.0&0.016&A0\\
2014.7288&56924& 0.118&-0.018&0.084&0.086&-12.1&0.058&K14b\\
2014.7288&56924& 0.160&-0.077&0.289&0.300&-14.9&0.066&K14a\\
2014.7288&56924& 0.071&-0.246&0.533&0.587&-24.8&0.269&K13\\
\hline
15 Nov 2014&&&&&&&&\\
2014.8740&56977&0.613& 0.000&0.000&0.000&0.0&0.024& A0\\
2014.8740&56977&0.057&-0.040&0.147&0.152&-15.4&0.060& K14b\\
2014.8740&56977&0.089&-0.094&0.336&0.349&-15.7&0.077& K14a\\
2014.8740&56977&0.025&-0.328&0.576&0.663&-29.7&0.226& K13\\
\hline
5 Dec 2014&&&&&&&&\\
2014.9288&56997& 0.655& 0.000&0.000&0.000&0.0&0.025&A0\\
2014.9288&56997& 0.092&-0.025&0.109&0.112&-12.9&0.065&K14b\\
2014.9288&56997& 0.105&-0.094&0.319&0.332&-16.4&0.069&K14a\\
2014.9288&56997& 0.046&-0.333&0.636&0.717&-27.6&0.366&K13\\
\hline
29 Dec 2014&&&&&&&&\\
2014.9945&57021& 0.664& 0.000&0.000&0.000&0.0&0.026&A0\\
2014.9945&57021& 0.079&-0.084&0.267&0.280&-17.5&0.105&K14b\\
2014.9945&57021& 0.114&-0.124&0.419&0.437&-16.5&0.115&K14a\\
2014.9945&57021& 0.038&-0.456&0.688&0.826&-33.5&0.587&K13\\
\hline
14 Feb 2015&&&&&&&&\\
2015.1233&57067&0.899& 0.000&0.000&0.000&0.0&0.021&A0\\
2015.1233&57067&0.070&-0.090&0.302&0.315&-16.5&0.123&K14a\\
2015.1233&57067&0.286&-0.158&0.463&0.489&-18.9&0.196&K14b\\
2015.1233&57067&0.031&-0.426&0.759&0.870&-29.3&0.420&K13\\
\hline
11 Apr 2015&&&&&&&&\\
2015.2767&57123&0.679& 0.000&0.000&0.000&0.0&0.018& A0\\
2015.2767&57123&0.119&-0.008&0.120&0.121& -3.9&0.048& K15\\
2015.2767&57123&0.111&-0.099&0.368&0.381&-15.0&0.110&K14b\\
2015.2767&57123&0.084&-0.156&0.533&0.555&-16.3&0.137&K14a\\
\hline
11 May 2015&&&&&&&&\\
2015.3589&57153& 0.354& 0.000&0.000&0.000&0.0&0.028& A0\\
2015.3589&57153& 0.103&-0.017&0.204&0.205& -4.7&0.040& K15\\
2015.3589&57153& 0.052&-0.121&0.388&0.407&-17.4&0.101& K14b\\
2015.3589&57153& 0.084&-0.177&0.568&0.595&-17.3&0.195& K14a\\
\hline
9 Jun 2015&&&&&&&&\\
2015.4385&57182&0.440& 0.000&0.000&0.000&0.0&0.016&A0\\
2015.4385&57182&0.121&-0.037&0.302&0.304& -6.9&0.049&K15\\
2015.4385&57182&0.050&-0.166&0.458&0.487&-19.9&0.112&K14b\\
2015.4385&57182&0.097&-0.232&0.634&0.675&-20.1&0.253&K14a\\
\hline
2 Jul 2015&&&&&&&&\\
2015.5014&57205&0.469& 0.000&0.000&0.000&0.0&0.014& A0\\
2015.5014&57205&0.092&-0.050&0.360&0.363& -8.0&0.051& K15\\
2015.5014&57205&0.059&-0.178&0.514&0.544&-19.1&0.176& K14b\\
2015.5014&57205&0.060&-0.269&0.651&0.704&-22.4&0.238& K14a\\
\hline
\hline
\end{tabular}
\end{table*}

\begin{table*}[!htbp]
\caption{Time evolution of polarization parameters (percentage and angle) for the core A0 observed from the jet of \s4}             
\label{table:CorePol_LC}
\centering                          
\begin{tabular}{c c c c c c c c c}        
\hline
\hline
MJD&  P$\pm$dP(\%)&EVPA$\pm$dE(deg)\\
\hline
56924&5.22$\pm$ 0.77&5.25$\pm$ 4.23\\
56977&6.99$\pm$0.80&  16.86$\pm$ 3.28\\
56997&7.74$\pm$0.72& -16.57$\pm$ 2.64\\
57021&8.15$\pm$0.69&  -7.33$\pm$ 2.43\\
57067&9.78$\pm$0.94&0.31$\pm$ 2.74\\
57123&8.52$\pm$0.41&  -7.03$\pm$ 1.37\\
57153&2.38$\pm$0.83&  -7.00$\pm$ 9.93\\
57182&3.19$\pm$0.63&  -9.34$\pm$ 5.66\\
57205&1.06$\pm$0.56& -51.79$\pm$15.3\\
\hline
\hline
\end{tabular}
\end{table*}

\section{Swift-XRT full dataset}\label{app:swiftxrt}
Table \ref{table:swiftXRT} collects all the analyzed exposures for the {\it Swift}-XRT dataset described in Section \ref{sec:mwlcoverage}. Fluxes have been extracted from a 20 pixel circular aperture. A different aperture was used on 2015 February 17 (MJD 57070.76), due to pile-up effects.
\begin{table*}[!htbp]
\caption{\s4 Swift-XRT exposures. For each observation, identified by its date and {\it Swift} observation identifier, we present: the duration of the exposure, the integrated energy flux in 2 energy bands, the best-fit spectral index, the $\chi^2$ and degrees of freedom of the fit.  }             
\label{table:swiftXRT}
\centering                          
\begin{tabular}{c c c c c c c c c}        
\hline
\hline
DATE-TIME&MJD&EXP&F(2-10 keV) &F(0.3-10 keV) &INDEX&$\chi^2_{\rm RED}$&DOF&OBSID\\
         &   &   & [10$^{\rm -12}$]& [10$^{\rm -12}$]&&&&\\
         &   & [s]  &  [erg cm$^{\rm -2}$ s$^{\rm -1}$] &  [erg cm$^{\rm -2}$ s$^{\rm -1}$] &&&&\\
\hline
&&&&&&&&\\
2006-07-04T00:49:40 & 53920.04 & 8620.6 & 2.76$^{\rm +0.22}_{\rm -0.19}$ & 4.08$^{\rm +0.20}_{\rm -0.21}$ & 1.62$\pm$0.06 & 0.69 & 30 & 00035381001 \\
2007-03-28T09:06:11 & 54187.38 & 3578.6 & 2.00$^{\rm +0.34}_{\rm -0.31}$ & 3.12$^{\rm +0.42}_{\rm -0.33}$ & 1.72$\pm$0.12 & 1.16 & 9 & 00036326001 \\
2008-01-10T01:09:39 & 54475.05 & 3748.5 & 1.61$^{\rm +0.23}_{\rm -0.23}$ & 2.68$^{\rm +0.29}_{\rm -0.26}$ & 1.82$\pm$0.11 & 1.11 & 10 & 00036326002 \\
2008-01-11T01:20:01 & 54476.06 & 2891.9 & 2.47$^{\rm +0.47}_{\rm -0.41}$ & 3.61$^{\rm +0.47}_{\rm -0.39}$ & 1.60$\pm$0.12 & 0.25 & 8 & 00036326003 \\
2008-01-15T16:10:28 & 54480.67 & 1513.4 & 3.71$^{\rm +1.68}_{\rm -1.14}$ & 4.84$^{\rm +1.53}_{\rm -0.96}$ & 1.34$\pm$0.23 & 0.89 & 3 & 00036326004 \\
2009-01-09T10:57:37 & 54840.46 & 10524.0 & 1.08$^{\rm +0.12}_{\rm -0.12}$ & 1.67$^{\rm +0.13}_{\rm -0.15}$ & 1.70$\pm$0.08 & 1.32 & 14 & 00036326005 \\
2009-11-01T22:49:53 & 55136.95 & 2784.5 & 1.81$^{\rm +0.42}_{\rm -0.27}$ & 2.48$^{\rm +0.37}_{\rm -0.38}$ & 1.46$\pm$0.16 & 0.17 & 4 & 00036326006 \\
2009-11-05T08:26:28 & 55140.35 & 2906.9 & 1.40$^{\rm +0.51}_{\rm -0.36}$ & 2.06$^{\rm +0.42}_{\rm -0.40}$ & 1.60$\pm$0.21 & 1.31 & 2 & 00036326007 \\
2009-12-12T18:45:25 & 55177.78 & 3848.3 & 3.72$^{\rm +0.39}_{\rm -0.38}$ & 4.95$^{\rm +0.35}_{\rm -0.33}$ & 1.39$\pm$0.08 & 0.89 & 15 & 00036326008 \\
2010-01-23T14:26:34 & 55219.60 & 8873.9 & 3.98$^{\rm +0.25}_{\rm -0.25}$ & 5.63$^{\rm +0.28}_{\rm -0.23}$ & 1.52$\pm$0.04 & 1.18 & 48 & 00090100001 \\
2010-03-12T05:57:53 & 55267.25 & 7980.6 & 3.47$^{\rm +0.22}_{\rm -0.27}$ & 4.89$^{\rm +0.19}_{\rm -0.24}$ & 1.52$\pm$0.05 & 1.69 & 38 & 00090100003 \\
2011-10-13T04:06:03 & 55847.17 & 1563.3 & 2.95$^{\rm +1.14}_{\rm -0.75}$ & 3.95$^{\rm +1.07}_{\rm -0.76}$ & 1.40$\pm$0.23 & 1.56 & 2 & 00036326009 \\
2011-10-14T13:19:43 & 55848.56 & 3074.2 & 1.67$^{\rm +0.29}_{\rm -0.35}$ & 2.49$^{\rm +0.38}_{\rm -0.31}$ & 1.64$\pm$0.14 & 1.76 & 6 & 00036326010 \\
2014-04-28T14:10:59 & 56775.59 & 1540.8 & 2.16$^{\rm +0.43}_{\rm -0.48}$ & 3.06$^{\rm +0.52}_{\rm -0.43}$ & 1.53$\pm$0.16 & 1.81 & 3 & 00091892001 \\
2014-05-28T20:08:46 & 56805.84 & 1920.4 & 5.36$^{\rm +0.74}_{\rm -0.65}$ & 7.40$^{\rm +0.64}_{\rm -0.58}$ & 1.48$\pm$0.10 & 1.47 & 12 & 00091892002 \\
2014-06-25T20:09:39 & 56833.84 & 1670.7 & 1.17$^{\rm +0.51}_{\rm -0.32}$ & 2.05$^{\rm +0.50}_{\rm -0.34}$ & 1.90$\pm$0.23 & 0.32 & 2 & 00091892003 \\
2014-11-17T23:06:57 & 56978.96 & 3262.1 & 12.08$^{\rm +0.95}_{\rm -0.80}$ & 15.07$^{\rm +0.74}_{\rm -0.81}$ & 1.20$\pm$0.06 & 1.21 & 32 & 00033530001 \\
2014-11-22T13:31:43 & 56983.56 & 4108.0 & 3.74$^{\rm +0.35}_{\rm -0.29}$ & 5.59$^{\rm +0.38}_{\rm -0.33}$ & 1.64$\pm$0.07 & 1.65 & 23 & 00033530002 \\
2015-01-27T19:19:19 & 57049.81 & 1942.9 & 3.89$^{\rm +0.49}_{\rm -0.45}$ & 6.02$^{\rm +0.59}_{\rm -0.56}$ & 1.70$\pm$0.10 & 1.31 & 10 & 00033530003 \\
2015-02-13T17:01:10 & 57066.71 & 1962.9 & 11.15$^{\rm +0.84}_{\rm -0.82}$ & 18.45$^{\rm +0.92}_{\rm -0.77}$ & 1.81$\pm$0.05 & 1.02 & 41 & 00033530004 \\
2015-02-15T07:15:48 & 57068.30 & 1893.0 & 10.35$^{\rm +0.91}_{\rm -0.84}$ & 14.38$^{\rm +0.84}_{\rm -0.84}$ & 1.49$\pm$0.07 & 1.02 & 24 & 00033530008 \\
2015-02-16T13:17:09 & 57069.55 & 1905.4 & 16.22$^{\rm +1.06}_{\rm -1.18}$ & 22.73$^{\rm +1.12}_{\rm -1.10}$ & 1.51$\pm$0.05 & 0.70 & 39 & 00033530009 \\
2015-02-17T18:20:49 & 57070.77 & 1775.6 & 21.32$^{\rm +1.42}_{\rm -1.47}$ & 31.82$^{\rm +1.67}_{\rm -1.31}$ & 1.64$\pm$0.05 & 1.03 & 36 & 00033530010 \\
2015-02-18T10:00:55 & 57071.42 & 2092.7 & 14.92$^{\rm +1.17}_{\rm -0.94}$ & 20.95$^{\rm +0.97}_{\rm -1.12}$ & 1.51$\pm$0.05 & 1.25 & 36 & 00033530011 \\
2015-02-19T08:22:52 & 57072.35 & 1071.3 & 13.59$^{\rm +1.23}_{\rm -1.43}$ & 19.35$^{\rm +1.41}_{\rm -1.30}$ & 1.54$\pm$0.08 & 0.81 & 18 & 00033530012 \\
2015-02-20T16:19:13 & 57073.68 & 983.9 & 10.89$^{\rm +1.38}_{\rm -0.90}$ & 15.92$^{\rm +1.59}_{\rm -1.16}$ & 1.60$\pm$0.09 & 0.41 & 14 & 00033530013 \\
2015-02-21T19:29:42 & 57074.81 & 1735.6 & 15.54$^{\rm +1.07}_{\rm -1.34}$ & 20.09$^{\rm +1.26}_{\rm -1.16}$ & 1.31$\pm$0.06 & 1.03 & 28 & 00033530014 \\
2015-02-22T14:41:28 & 57075.61 & 1937.9 & 14.28$^{\rm +1.21}_{\rm -1.05}$ & 18.66$^{\rm +1.20}_{\rm -0.95}$ & 1.34$\pm$0.05 & 1.37 & 30 & 00033530015 \\
2015-02-23T03:29:19 & 57076.15 & 994.0 & 12.62$^{\rm +1.59}_{\rm -1.31}$ & 16.81$^{\rm +1.65}_{\rm -1.47}$ & 1.39$\pm$0.09 & 0.93 & 13 & 00033530017 \\
2015-02-24T05:07:19 & 57077.21 & 1371.0 & 22.37$^{\rm +1.40}_{\rm -1.70}$ & 27.47$^{\rm +2.16}_{\rm -1.78}$ & 1.15$\pm$0.06 & 1.05 & 27 & 00033530018 \\
2015-03-04T19:34:43 & 57085.82 & 2205.1 & 12.77$^{\rm +0.83}_{\rm -0.72}$ & 18.42$^{\rm +0.86}_{\rm -0.85}$ & 1.57$\pm$0.05 & 1.13 & 40 & 00033530019 \\
2015-03-05T06:44:02 & 57086.28 & 1578.3 & 16.54$^{\rm +1.55}_{\rm -1.13}$ & 20.93$^{\rm +1.53}_{\rm -1.23}$ & 1.25$\pm$0.06 & 1.03 & 24 & 00033530020 \\
2015-03-06T11:26:12 & 57087.48 & 1875.5 & 7.93$^{\rm +0.83}_{\rm -0.83}$ & 10.88$^{\rm +0.85}_{\rm -1.01}$ & 1.46$\pm$0.08 & 0.96 & 18 & 00033530021 \\
2015-03-07T10:04:13 & 57088.42 & 1311.1 & 9.30$^{\rm +1.19}_{\rm -1.18}$ & 14.04$^{\rm +1.22}_{\rm -1.00}$ & 1.66$\pm$0.10 & 0.99 & 13 & 00033530022 \\
2015-03-08T14:43:10 & 57089.61 & 1210.7 & 8.69$^{\rm +1.30}_{\rm -1.82}$ & 11.41$^{\rm +1.46}_{\rm -1.42}$ & 1.35$\pm$0.13 & 1.13 & 6 & 00033530023 \\
2015-03-09T16:02:02 & 57090.67 & 1838.0 & 8.47$^{\rm +0.99}_{\rm -0.95}$ & 10.76$^{\rm +0.96}_{\rm -0.78}$ & 1.26$\pm$0.08 & 0.67 & 14 & 00033530024 \\
2015-03-10T06:27:07 & 57091.27 & 1098.8 & 7.19$^{\rm +1.42}_{\rm -1.11}$ & 9.27$^{\rm +1.41}_{\rm -1.19}$ & 1.30$\pm$0.13 & 1.00 & 6 & 00033530025 \\
2015-03-11T06:22:22 & 57092.27 & 1863.0 & 8.66$^{\rm +0.83}_{\rm -0.90}$ & 11.49$^{\rm +0.83}_{\rm -0.83}$ & 1.38$\pm$0.07 & 0.48 & 16 & 00033530026 \\
2015-06-21T17:29:18 & 57194.73 & 1465.9 & 5.18$^{\rm +1.88}_{\rm -1.28}$ & 6.38$^{\rm +1.99}_{\rm -1.33}$ & 1.16$\pm$0.24 & 0.70 & 1 & 00033829001 \\
2015-06-22T20:25:32 & 57195.85 & 1965.4 & 3.78$^{\rm +0.78}_{\rm -0.51}$ & 5.12$^{\rm +0.58}_{\rm -0.55}$ & 1.43$\pm$0.13 & 0.80 & 6 & 00033829002 \\
2015-06-24T04:33:29 & 57197.19 & 2202.6 & 2.82$^{\rm +0.56}_{\rm -0.44}$ & 4.06$^{\rm +0.61}_{\rm -0.49}$ & 1.57$\pm$0.15 & 0.72 & 6 & 00033829004 \\
2015-06-25T02:44:56 & 57198.12 & 1635.7 & 4.07$^{\rm +1.25}_{\rm -1.09}$ & 5.30$^{\rm +1.26}_{\rm -1.01}$ & 1.33$\pm$0.19 & 0.54 & 4 & 00033829005 \\
2015-06-26T01:06:48 & 57199.05 & 986.4 & 4.14$^{\rm +1.16}_{\rm -1.01}$ & 5.83$^{\rm +1.49}_{\rm -0.97}$ & 1.52$\pm$0.20 & 1.04 & 3 & 00033829006 \\
2015-06-27T04:36:46 & 57200.19 & 1808.0 & 4.19$^{\rm +0.74}_{\rm -0.55}$ & 5.70$^{\rm +0.86}_{\rm -0.72}$ & 1.44$\pm$0.12 & 1.10 & 7 & 00033829007 \\
2015-06-28T00:54:48 & 57201.04 & 1773.1 & 6.49$^{\rm +1.57}_{\rm -1.51}$ & 7.62$^{\rm +1.78}_{\rm -1.57}$ & 0.99$\pm$0.19 & 0.34 & 4 & 00033829008 \\
2015-06-29T05:54:31 & 57202.25 & 1748.1 & 3.03$^{\rm +0.55}_{\rm -0.46}$ & 4.40$^{\rm +0.59}_{\rm -0.59}$ & 1.58$\pm$0.13 & 1.20 & 6 & 00033829009 \\
2015-06-30T09:00:28 & 57203.38 & 1962.9 & 3.11$^{\rm +0.57}_{\rm -0.61}$ & 4.21$^{\rm +0.76}_{\rm -0.56}$ & 1.43$\pm$0.15 & 0.76 & 5 & 00033829010 \\
&&&&&&&&\\

\hline
\hline
\end{tabular}
\end{table*}


\end{appendix}

\end{document}